\RequirePackage{fix-cm}
\documentclass[smallextended]{svjour3}       % onecolumn (second format)
\smartqed  % flush right qed marks, e.g. at end of proof
\usepackage{graphicx}
\DeclareGraphicsExtensions{.bmp}

\newcommand{\be}{\begin{equation}}
\newcommand{\ee}{\end{equation}}
\newcommand{\fr}{\frac}

\newcommand{\iz}{\left}
\newcommand{\de}{\right}

\newcommand{\tb}{\textbf}
\newcommand{\bit}{\begin{itemize}}
	\newcommand{\eit}{\end{itemize}}

\newcommand{\bra}[1]{\langle #1|}
\newcommand{\ket}[1]{|#1\rangle}

\begin{document}

\title{Quantum confinement effects on the near field enhancement in metallic nanoparticles%\thanks{Grants or other notes
}


\author{Mario Zapata-Herrera        \and 
        Jefferson Fl\'orez     \and  
        Angela S. Camacho    \and  
        Hanz Y. Ram\'irez 
}


\institute{M. Zapata-Herrera   \at
            Departamento de F\'isica, Universidad de los Andes, Bogot\'a, D.C. 111711, Colombia \\
             \emph{Present address:} Materials Physics Center CSIC-UPV/EHU and Donostia International Physics Center DIPC, Donostia-San Sebasti\'an 20018, Spain.\\              
               \and
              Jefferson Fl\'orez  \at
              Physics Department, University of Ottawa, Ottawa ON  K1N 6N5, Canada. \\
              \and
              Angela S. Camacho \at
              Departamento de F\'isica, Universidad de los Andes, Bogot\'a, D.C. 111711, Colombia \\
              \and
               Hanz Y. Ram\'irez  \at
               Grupo de F\'isica Te\'orica y Computacional, Escuela de F\'isica, Universidad Pedag\'ogica y Tecnol\'ogica de Colombia (UPTC), Tunja 150003, Boyac\'a, Colombia. \\
              \email{hanz.ramirez@uptc.edu.co}
}            
               
%              Tel.: +123-45-678910\\
%              Fax: +123-45-678910\\
%              \email{fauthor@example.com}           %  \\
%%             \emph{Present address:} of F. Author  %  if needed

\date{Received: date / Accepted: date}

\maketitle

\begin{abstract}
In this work we study the strong confinement effects on the electromagnetic response of metallic nanoparticles. We calculate the field enhancement factor for nanospheres of various radii by using optical constants obtained from both classical and quantum approaches, and compare their size dependent features. To evaluate the scattered near field, we solve the electromagnetic wave equation within a finite element framework. When quantization of electronic states is considered for the input optical functions, a significant blue-shift in the resonance of the enhanced field is observed, in contrast to the case in which functions obtained classically are used. Furthermore, a noticeable underestimation of the field amplification is found in the calculation based on a classical dielectric function.     
Our results are in good agreement with available experimental reports and provide relevant information on the cross-over between classical and quantum regime, useful in potentiating nanoplasmonics applications. 
\end{abstract}

\section{Introduction}
\label{intro}
A collective resonant response to stimulating electromagnetic fields by the conduction electrons in metallic nanoparticles, is widely known as localized surface plasmon resonance (LSPR), whose characteristic peak depends on geometrical and compositional factors, as well as on the nature of the incident light (i.e. polarization, intensity and wavelength) \cite{Wei,Murray,Carsten,Rycenga,Igarachi,Li}. 

While for bulky nanoparticles (tens of nanometers or more in typical length), behavior of the LSPR is well understood in terms of the classical Drude's model, recently reported blue-shifts of the LSPR in particles with radius below 10 nanometers, reveals the need for a quantum treatment of the involved carriers in order to more accurately describe the collective response in such strongly confined systems \cite{Lerme,Scholl,Garcia,Zuloaga,Marinica,Raza,Monreal}.

Understanding of the crossing between classical and quantum regimes and its effects on the optical response in this kind of systems, in addition to be appealing since the fundamental point of view, is becoming more and more relevant because of growing interest in nanoplasmonics applications.  

Plasmonic resonances in isolated nanoparticles under polarized electromagnetic radiation, produce a strong scattered electric field as compared to the intensity of the incident field \cite{Nordlander,Davoyan,Huang,zhuzhao}. This effect is promising for technological applications given that augmented fields open doors to more efficient scenarios for strong radiation-matter coupling in low dimensional semiconductor structures \cite{Fung,Singh,Ali,Anton}, and for magnified non-linearities in polar materials \cite{Ciraci,Zhu,Zapata}. 

In this work, we study such an enhancement effect by using classical and quantum models in obtaining the dielectric functions. This allows us to clearly compare the optical response from both approaches. 

The paper is organized as follows: In the first part the electromagnetic problem is described in terms of the dielectric properties of the nanoparticle. In the second part those dielectric functions are calculated in the framework of the classical Drude's model and also within a model that includes electron energy discretization from quantum confinement. In the last part, the near enhanced fields obtained from both approaches are compared, and a summary and conclusions are provided.  

\section{Field enhancement factor}
\label{sec:1}
 The electric field scattered by the nanosphere must satisfy the Maxwell equations in matter. This leads to the Helmholtz wave equation for the electric field vector $\vec{E}$, which in terms of the complex relative permitivity $\epsilon_r$, reads 
 
 \begin{equation}
\nabla \times \mu_r\iz(\nabla \times \vec{E}\de)-k_0^2\iz(\epsilon_r - \frac{i\sigma}{\omega\epsilon_0}\de)\vec{E}=0 \hspace*{1ex} ,
 \label{Maxwell}
 \end{equation}
 
where $\mu_r$ is the relative permeability (taken as 1 for metallic systems),  $\sigma$ the nanoparticle conductivity,  and $k_0=\omega \sqrt{\epsilon_0\mu_0}=\fr{\omega}{c_0}$ (with $c_0$, $\epsilon_0$ and $\mu_0$ the light velocity, vacuum permeability, and vacuum permitivity, respectively) is the incident light's wave number. \\

The dielectric function is required as input parameter to solve equation (\ref{Maxwell}) , and that function in turn depends on the frequency of the incident electromagnetic wave $\epsilon_r(\omega) = \epsilon_1(\omega) + i \epsilon_2(\omega)$. 

Once this input function is established, the vectorial wave equation (\ref{Maxwell}) can in principle be solved, yet in most cases numerical treatment is necessary \cite{Chang}. In this case, we solve computationally this complex equation by means of a standard finite element method \cite{Arnaud,Misic,Santana,forou,Jacak1,comsol}. 

We define the field enhancement factor ($FEF$) as the squared norm of the ratio between the scattered electric field $\vec{E}_{out}$ in any point of the nanosphere surrounding (near field), and the amplitude of the incident $z$-polarized electric field $\vec{E}_{inc}$

\be
FEF=\fr{|\vec{E}_{out}|^2}{|\vec{E}_{inc}|^2} \hspace*{1ex} .
\ee

\begin{figure*}
		\begin{center}
			\includegraphics[scale=0.45]{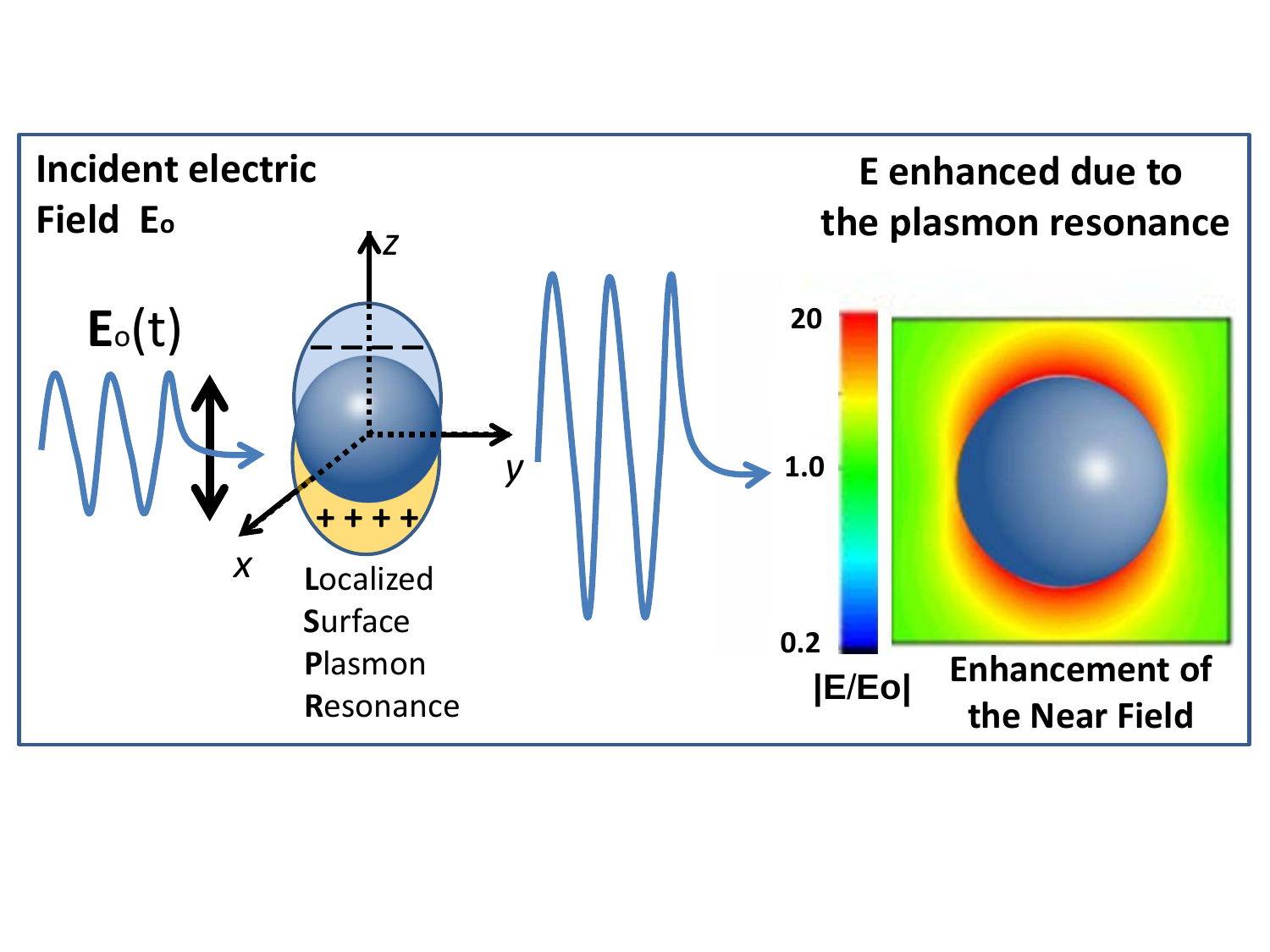}
			\caption{Schematics of the stimulated plasmonic polarization in the nanosphere and the corresponding electric field modification in its surrounding.}
			\label{EF}
		\end{center}
\end{figure*}

The studied system is represented in figure (\ref{EF}), where the polarization effect of the incident wave on the conduction electronic plasma in the nanosphere, and the associated electric field modification are depicted. 

% Optical parameters $\eta$ and $\kappa$ as functions of $\omega$, are related to  the real  and imaginary parts of the complex dielectric function [$\epsilon_r(\omega) = \epsilon_1(\omega) + i \epsilon_2(\omega)$], by \cite{etakappa}
%
%\be
%\eta(\omega)=\sqrt{\fr{\epsilon_1(\omega)}{2}+\fr{1}{2}\sqrt{\epsilon_1(\omega)^2+\epsilon_2(\omega)^2}} \hspace*{1ex} ,
%\ee
%
%and
%
%\be
%\kappa(\omega)=\fr{\epsilon_2(\omega)}{2\sqrt{\fr{\epsilon_1(\omega)}{2}+\fr{1}{2}\sqrt{\epsilon_1(\omega)^2+\epsilon_2(\omega)^2}}} \hspace*{1ex} .
%\ee
%
% Thus, the local field is obtained from solving eq. (\ref{Maxwell}) and the FEF is then directly calculated.

%%%%%%%%%%%%%%%%%%%%%%%%%%%%%%%%%%%%%%%%%%%%%%%%%%%%%%%%%%%%
\section{Dielectric optical response}
%%%%%%%%%%%%%%%%%%%%%%%%%%%%%%%%%%%%%%%%%%%%%%%%%%%%%%%%%%%%

\begin{figure}
	\begin{center}
		\includegraphics[scale=0.45]{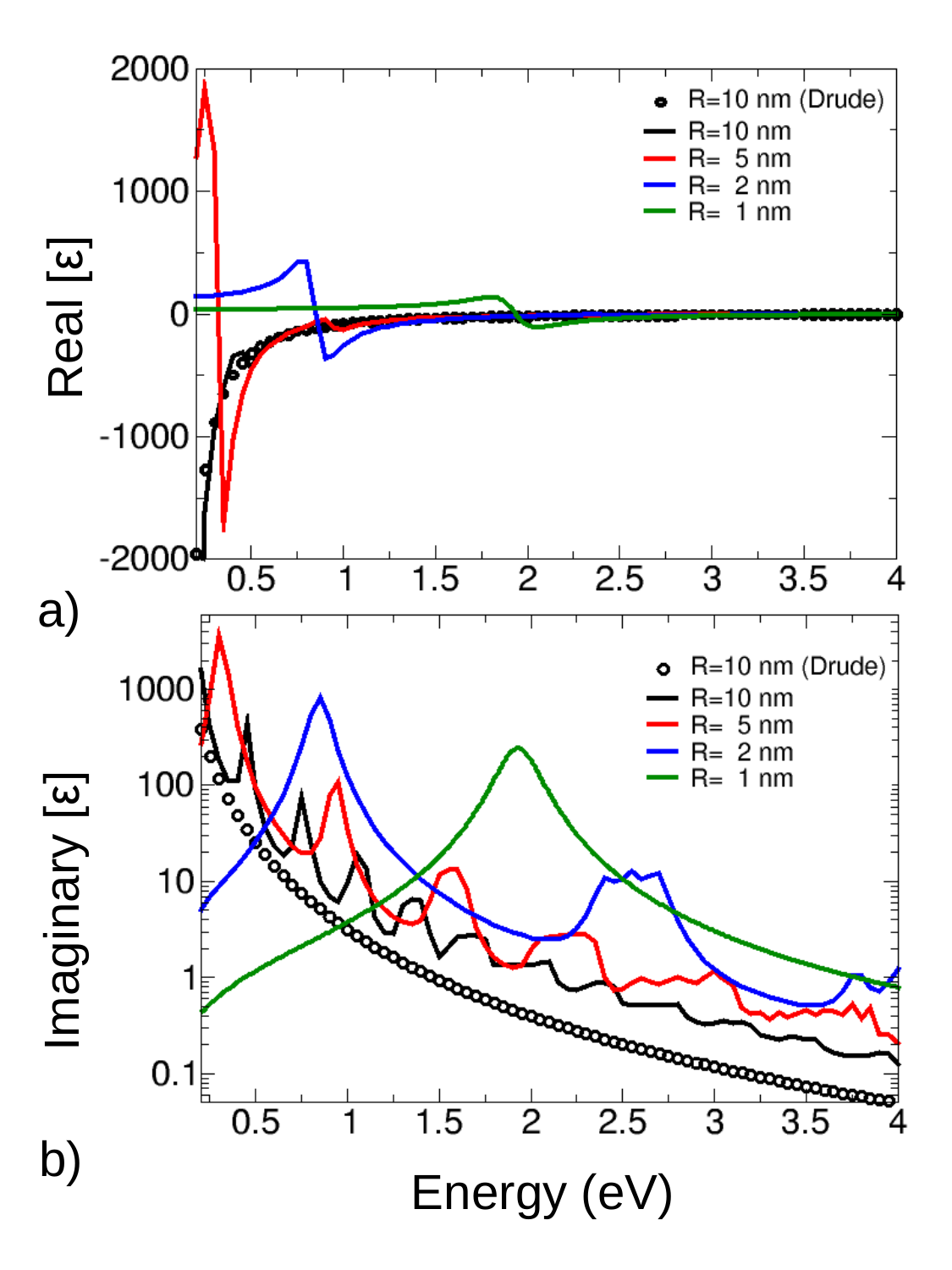}
		\caption{(a) Real part of the relative permitivity $\epsilon_1(\omega)$, obtained from both of the studied approaches. (b) Imaginary part of the relative permitivity $\epsilon_2(\omega)$, obtained from both of the studied approaches.}
		\label{dielectric}
	\end{center}
\end{figure}

%%%%%%%%%%%%%%%%%%%%%%%%%%%%%%%%%%%%%%%%%%%%%
\subsection{Classical model}
%%%%%%%%%%%%%%%%%%%%%%%%%%%%%%%%%%%%%%%%%%%%%

The dielectric function of a metal in a classical framework, can be described by a simplified model where the conduction electrons are considered to constitute a gas with a number $n$ of charges per volume unit, moving on a background of positive ion cores. Based on a classical motion equation for a carrier in a plasma under the influence of an external electric field and assuming a harmonic response, the dielectric function becomes \cite{Zuloaga,Marinica},

\be
\epsilon(\omega)=\epsilon_\infty - \fr{\omega_p^2}{\omega^2+i\gamma\omega} \hspace*{1ex} ,
\label{classic}
\ee

where $\epsilon_{\infty}$ corresponds to the interband screening contribution from the core electrons, while the term $\gamma=1/\tau$  is a damping frequency related to elastic collisions and depends on the relaxation time of the free electron gas $\tau$.  For its part, the plasmon frequency

\be
\omega_p^2=\fr{4\pi ne^2}{m^*} \hspace*{1ex} ,
\ee   

depends on the electronic density $n$, the elemental electric charge $e$, and the electron effective mass $m^*$ \cite{drudemodel}.

These underlaying assumptions show how this so-called Drude model considers the relevant electrons as completely classical particles. However, in despite of such a limitation, that approach has been regularly used to describe plasmonic behavior in nanoparticles \cite{Kelly,Grady,Noguez,Esteban,Muneton}.

%%%%%%%%%%%%%%%%%%%%%%%%%%%%%%%%%%%%%%%%%%%%%
\subsection{Quantum model}
%%%%%%%%%%%%%%%%%%%%%%%%%%%%%%%%%%%%%%%%%%%%%

Along with the size reduction of the metallic particles, the ``continuous'' conduction band of the nanostructure is expected to break up into well discretized states \cite{Genzel,Garcia}. 

As a first attempt to describe the optical response of the electron gas in a metal nanoparticle, Genzel et al. in Ref. \cite{Genzel} presented a quantum model where electrons in the conduction band remained non-interacting, but experienced confinement by a hard-wall potential. Within this model, the dielectric function is given by 

\be
\epsilon_r (\omega)=\epsilon_{\infty}+\fr{\omega_p^2}{N}\sum_{i,f}\fr{s_{if}(F_i-F_f)}{\omega_{if}^2-\omega^2-i\omega \gamma_{if}} \hspace*{1ex} ,
\label{quantum}
\ee

where $N$ the total number of valence electrons in the nanoparticle and $\epsilon_{\infty}$ is the same as in the classical case.  $s_{if}$, $\omega_{if} \equiv \frac{E_f - E_i}{\hbar}$, and $\gamma_{if}$, are the oscillator strength, frequency, and damping for the dipole transition from an initial state $\mid i \rangle$ to a final state $ \mid f \rangle$, respectively ($F_i$ and $F_f$  account for the corresponding values of the Fermi-Dirac distribution function).

In particular, the oscillator strength terms are the components that account for the effects of energy discretization, and include the geometrical features of the confinement. They are related to the dipole moment in the polarization direction (which in this study is chosen to be $z$) \cite{Pccp}, according to  

\be
s_{if}=\fr{2 m^* \omega_{if}}{\hbar}|\langle f|z|i \rangle |^2 \hspace*{1ex}.
\label{strenght}
\ee

Following the geometry of the nanoparticles, we consider the infinite spherical potential well, as done in reference \cite{Scholl}. However we go beyond the asymptotic approximation used in that work, by taking the full solutions of the Schr\"odinger equation for such a confining potential, i.e.  wave functions of the form 

\begin{equation}
	\psi_{n,l,m}(r,\theta,\phi)=\fr{1}{|j_{l+1}(\alpha_{nl})|} \sqrt{ \fr{2}{R^3}}  j_l \left( \fr{\alpha_{nl}}{R}r \right)Y_l^m(\theta,\phi) \hspace*{1ex},
	\label{Psi}
\end{equation}

where $j_l$ represents the $l$-th spherical Bessel function, $Y_l^m$ the standard spherical harmonics, and $\alpha_{nl}$ is the $n$-th zero of $j_l$ (i.e. $j_l(\alpha_{nl})=0$ for $n=0,1,2,\dots$) \cite{Arfken}.

Correspondingly, the discretized eigenenergies $E_{n,l}$ are related to the zeros of the spherical Bessel functions by the expression 

\begin{equation}
E_{n,l}=\frac{\hbar^2\alpha_{nl}^2}{2mR^2} \hspace*{1ex}.
\label{E}
\end{equation}

By using $z=r \cos\theta$ , and the wave functions from equation (\ref{Psi}), the dipole moment in equation (\ref{strenght}) can be obtained through the integral 
 
\begin{equation}
	|\bra{f}z\ket{i}|=\int_{0}^{2\pi}\int_{0}^{\pi}\int_{0}^{R}r^2\sin\theta dr d\theta d\phi\Psi_{n_f,l_f,m_f}^*(r,\theta,\phi)r \cos \theta \Psi_{n_i,l_i,m_i}(r,\theta,\phi) \hspace*{1ex},
\end{equation}

whose angular and radial parts become, respectively

\begin{eqnarray}
	I_{ang}&=&\sqrt{\fr{(l_i+m_i+1)(l_i-m_i+1)}{(2l_i+1)(2l_i+3)}}\delta_{\Delta l,+1}+\sqrt{\fr{(l_i+m_i)(l_i-m_i)}{(2l_i+1)(2l_i-1)}}\delta_{\Delta l,-1} \hspace*{1ex},
	\nonumber\\
	I_{rad}&=&\fr{1}{|j_{l_f+1}(\alpha_{n_fl_f})|} \fr{1}{|j_{l_i+1}(\alpha_{n_il_i})|}\left( \fr{2}{R^3}\right)\int_{0}^{R}dr  j_{l_f} \left( \fr{\alpha_{n_fl_f}}{R}r \right)r^3  j_{l_i} \left( \fr{\alpha_{n_il_i}}{R}r \right) \hspace*{1ex}. \nonumber \\
\end{eqnarray}

Thus, the integral reduces to

\begin{equation}
	|\bra{f}z\ket{i}|=\fr{I_{ang}}{|j_{l_f+1}(\alpha_{n_fl_f})||j_{l_i+1}(\alpha_{n_il_i})|}\left( \fr{2}{R^3}\right)\int_{0}^{R}dr  j_{l_f} \left( \fr{\alpha_{n_fl_f}}{R}r \right)r^3  j_{l_i} \left( \fr{\alpha_{n_il_i}}{R}r \right) \hspace*{1ex} ,
	\label{integral}
\end{equation}

which is different of zero only for values that satisfy $\Delta l=l_f-l_i=\pm 1$.

Equations (\ref{E}) and (\ref{integral}) allow for calculating the $s_{if}$ terms, conducting to obtain via equation (\ref{quantum}) a dielectric function that incorporates quantum features originated in size reduction.

%%%%%%%%%%%%%%%%%%%%%%%%%%%%%%%%%%%%%%%%%%%%%
\section{Results}
%%%%%%%%%%%%%%%%%%%%%%%%%%%%%%%%%%%%%%%%%%%%%

\begin{figure}
	\begin{center}
		\includegraphics[scale=0.45]{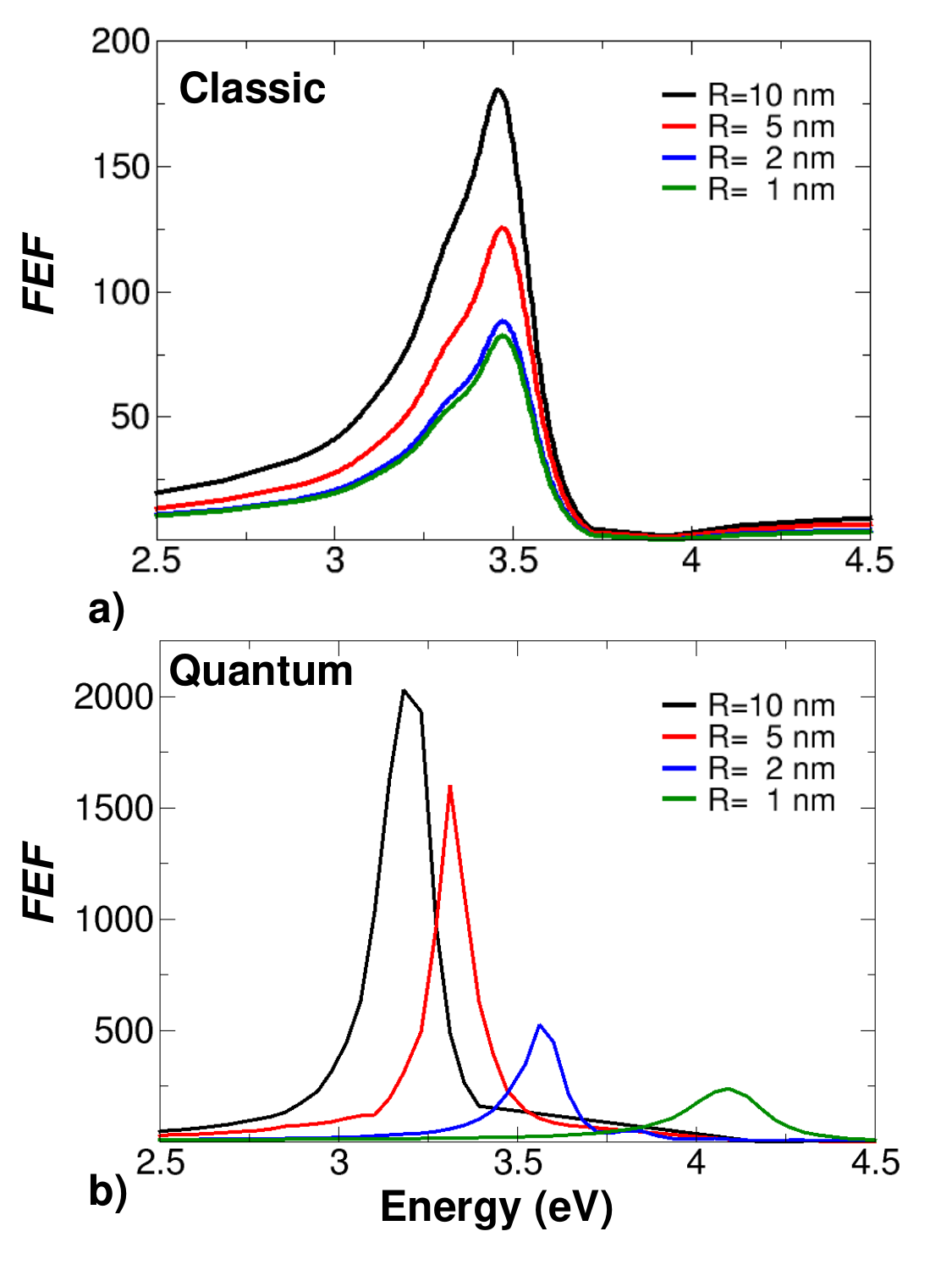}
		\caption{ (a) Field enhancement factor for silver nanospheres embedded in air, where $\epsilon_r(\omega)$ from the classical approach is used. (b) As in (a) but from the quantum approach.}
		\label{FEF}
	\end{center}
\end{figure}

For silver isolated nanoparticles of radii ranging from 1 to 10 nm, we compute the real and imaginary parts of the dielectric function $\epsilon_r (\omega)$, and input it in equation (\ref{Maxwell}) so that the $FEF$ can be obtained. 

In solving numerically the Helmholtz equation, the nanospheres were divided in a mesh of size decreasing tetrahedral elements, until convergence was achieved. The dielectric function within each considered model was calculated in frequency steps of $10^{13}$ Hz.

We carry out such a calculation in both frameworks, classical and quantum, by using equations (\ref{classic}) and (\ref{quantum}), respectively.

Within the classical approach, we take bulk optical constants from experiments by Johnson and Christie \cite{JandC}; whereas for the quantum approaches we use material parameters from Scholl et al. \cite{Scholl}.  

Figure \ref{dielectric}(a) [\ref{dielectric}(b)] shows the real [imaginary] part of the dielectric function obtained from each of the considered approaches.

%Besides the obvious fact that the dielectric function is size independent in the classical framework [equation (\ref{classic})], 

There are two remarkable differences between the dielectric functions obtained through these two approaches: First, the appearance in the quantum case of a fine structure due to inclusion of multiple transitions between discretized states. Second, also for the quantum case, the blue-shift of the main resonance with size-reduction \cite{Genzel}, which can be understood in terms of the increased energy separation between eigenstates when the confinement is stronger [the fact that there is only one dielectric function in the classical framework is directly related to the size insensitivity of equation (\ref{classic})].

Figures \ref{FEF}(a) and \ref{FEF}(b) present the obtained values for the field enhancement factor computed right on the north pole of the spherical nanoparticle, as calculated for the dielectric functions showed in figures \ref{dielectric}(a) and \ref{dielectric}(b). There, it can be observed how the magnitude and activation energy of the $FEF$ calculated by either the classical or the quantum approach, exhibit eye-catching differences. First, size dependence of the main plasmonic resonance in the quantum case appears in strong contrast to the stable peak position observed in the classical case. Second, an evident discrepancy in the intensity of the field enhancement is found.

Clearly, the insertion of quantum confinement effects in the dielectric function underlies these dissimilarities. 

On one side, the size dependent resonance is straightforwardly related to the enlargement of the energy separation between discretized states with increasing confinement, which is in complete agreement with results of absorption spectroscopy in colloids of silver nanoparticles \cite{charle}, and EELS experiments on single metal nanoparticles \cite{Scholl}. On the other hand, the underestimation caused by using the classical approach, should be associated to the fact of averaging the plasma contribution along all the spheric volume of the nanoparticle, while the quantum approach includes the wave function distribution, that concentrates close to the surface for excited states [the ones contributing the most to the summation in equation (\ref{quantum})] \cite{Monreal}.

Moderate underestimation of the $FEF$ obtained within the Drude model is in agreement with time dependent DFT calculations reported by Negre et al. in \cite{Negre}, but opposite to results from Zuloaga et al. in \cite{Zuloaga}. We would like to point out that the later work considers a damping constant around one order of magnitude larger than the one used in our calculations \cite{Blaber,Scholl}, increasing so the imaginary part of the dielectric function and inflating the collective response in the nanoparticle.      

Table \ref{percentage} shows the relative underestimation of the enhancement factor in the classical framework ($FEF_C$) as compared to that in the quantum framework ($FEF_Q$). 

%(by using data extracted from figure \ref{FEF}).     

\begin{table}
	\begin{center}
		\begin{tabular}{|c |c |c |c|c|}\hline
			R(nm)& 1 & 2 & 5 & 10 \\\hline
			$FEF_Q/FEF_C$ & 2.9 & 5.9 & 12.7 & 11.2  \\\hline
		\end{tabular}
	\end{center}
	\caption{Relative underestimation for the different studied nanospheres.}
	\label{percentage}
\end{table}

It is noteworthy that the proportional difference in the magnitude of $FEF$ for the compared approaches is neither maximum for the smallest nor the largest studied nanoparticles. This is important because it should be expected that at some radius larger than 10 nm, corrections associated with confinement effects start becoming negligible. Such non-monotonic behavior is here explicated in terms of a competition between the confinement effect (that is related to the surface to volume ratio and scales as $\frac{1}{r}$), and the number of available carriers constituting the stimulated plasmons (that is proportional to the number of atoms, accounts for the increase of $FEF$ with the nanosphere size in both approaches, and scales as $r^3$).    

As for the resonance frequency, one could at principle expect that for the largest simulated sphere the peak position would recover the classical limit, which is not observed in figure \ref{FEF}. It is important to mention that in the regime of ``large" nanospheres (R=10-100 nm), some other effects as dynamical surface screening and Lorentz friction become relevant \cite{Monreal,Jacak,Jacak1}, and their interplay alongside with softened quantum confinement contribute to a non-monotonic behavior for the plasmon resonance, before the classical limit is actually reached in micro-metric particles \cite{Scholl,Marzan}.       

\begin{figure}
		\begin{center}
			\includegraphics[scale=0.4]{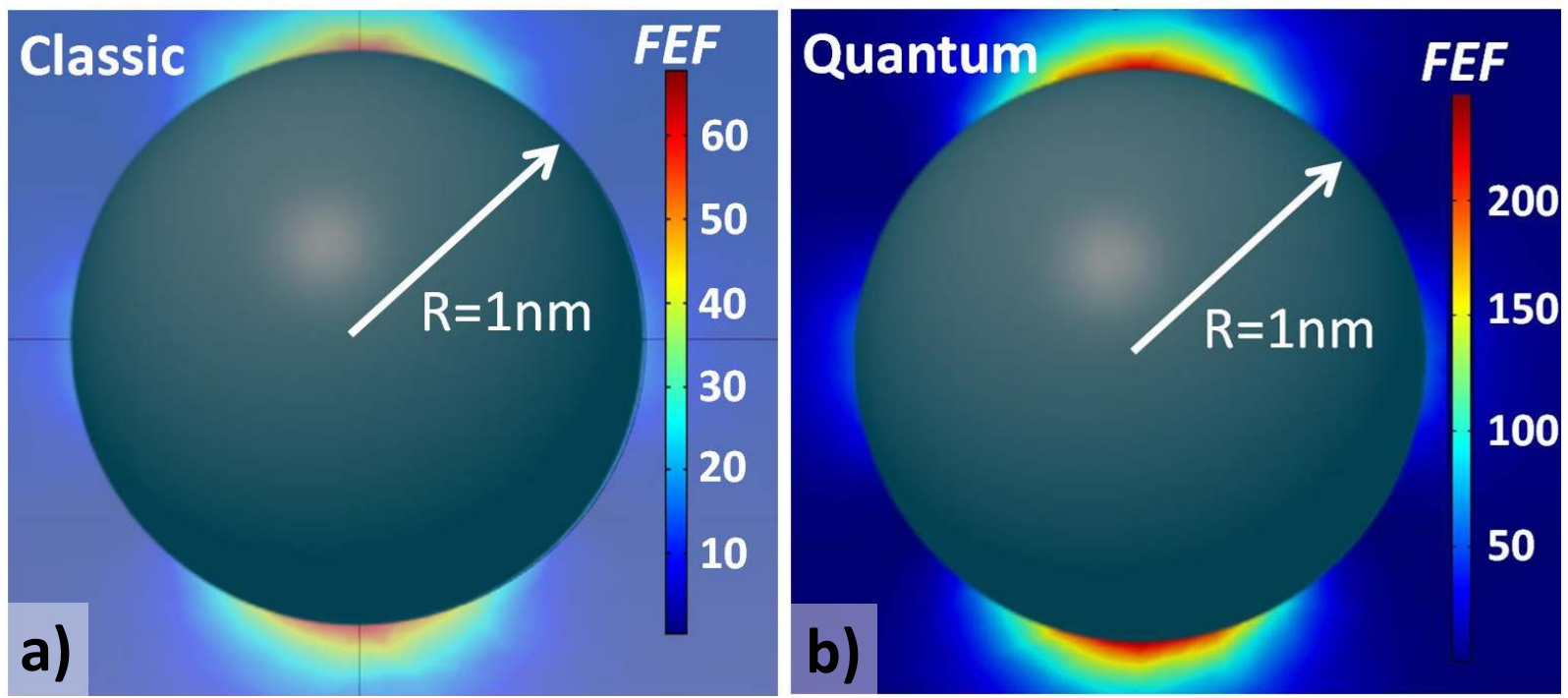}
			\caption{ Distribution of the near field enhancement factor in the nanosphere´s vicinity. (a) Obtained with the classical dielectric function, and (b) obtained with the quantum dielectric function.}
			\label{2D}
		\end{center}
\end{figure}

Figures \ref{2D}(a) and \ref{2D}(b) present the squared scattered fields normalized to the magnitude of the incident field at the vicinity of the nanosphere of radius $R=1$ nm, as obtained from the classical and quantum approaches, respectively. The shown 2D distributions are calculated in the corresponding resonance frequencies [as extracted from figures 3(a) and 3(b)].   

These graphs are consistent with previous works in which the dipole plasmon mode is found to be by far the main contribution to the field enhancement in symmetric structures \cite{Zhang,ZhangY}. Additionally they help to visualize how the LSPR is strengthened within a model in which the wave functions of the carriers constituting the plasma are considered.  

It is pertinent to mention that the general outline of our calculation is suitable for other geometries, as long as an appropriate confining potential is used. However, for arbitrary particle shapes, to deal with hundreds of numerical solutions of the eigenvalue problem for the summation in equation (\ref{quantum}), might result a challenge. 

Finally we would like to address the fact that in our model, for the sake of simplification, an infinite confinement potential is considered implying that the widely discussed spill-out effect \cite{Heer,Brack,Monreal}, is neglected. We would expect no qualitative change in our results if a finite potential was used, given that the physics underlying the resonance blue-shift and the classical underestimation of the $FEF$ is not particularly sensitive to the potential height \cite{Genzel}. However, to establish how significant would be the quantitative change in case the potential allows for non vanishing electron density beyond the sphere edge, constitutes an interesting extension of this work.

%In terms of the surface to volume ratio ($S/V$), it is noteworthy the pronounced change in this quantity for spheres between 1 and 10 nm in radius, as shown in table \ref{alpha}. The $S/V$ ratio increases from $6\%$ ($R=10$nm) to $60\%$ ($ R=1$ nm), and doubles between $R=2$ nm and $R=1$ nm this ratio doubles. Considering the surface effects, it becomes instinctive that for the biggest studied nanospheres, the results based on the quantum approach resemble the ones obtained through the classical approach. 
%
%
%
%\begin{table}[H]
%	\begin{center}
%		\begin{tabular}{|c |c |c |c|c |c | c| c|}\hline
%			R(nm)& $1nm$& $2nm$ & $5nm$ & $10nm$ & $12nm$ &$15nm$ &$20nm$\\\hline
%			$S/V$ (nm$^{-1}$)&$3$& $1.5$ &$0.6$ & $0.3$ & $0.25$ &$0.2$ &$0.15$ \\\hline
%		\end{tabular}
%	\end{center}
%	\caption{Surface to Volume ratio percentage vs. sphere radius.}
%	\label{alpha}
%\end{table}

%%%%%%%%%%%%%%%%%%%%%%%%%%%%%%%%%%%%%%%%%%%%%
\section{Summary and conclusions}
%%%%%%%%%%%%%%%%%%%%%%%%%%%%%%%%%%%%%%%%%%%%%

We studied numerically the effects of quantum confinement on the field enhancement factor of metallic nanoparticles stimulated by time dependent electromagnetic fields of varying frequencies. Remarkable differences in this optical response when calculated within a classical and a quantum framework, were observed.

Influence of the surface to volume ratio and total number of conduction carriers on the collective excitation has been illustrated, and the need for including quantum phenomena in correctly describing the properties of small metallic nanoparticles is evidenced.

According to our results, significant underestimation in magnitude, and spurious insensibility of the plasmonic-induced resonance to the nanoparticle size; should be expected if a classical approach is used to study the field augment in spheres with radii at the order of few nanometers.

Furthermore, the fact that the discrepancies found in the field enhancement were originated in the dielectric function used for each approach, suggest that alike effects could be anticipated in diverse geometries and in other optical responses (e.g. higher harmonic generation and frequency filtering). 

%It has been also shown that an increment in the refraction index of the nanoparticle embedding material, favors the field enhancement.

We trust these findings will contribute to the optimization of ensembles for field detection and/or measurements, and somehow advance the understanding on the cross-over between classical and quantum regimes in nanostructures.

\begin{acknowledgements}
The authors acknowledge the Department of Physics of the Universidad de Los Andes, and the UPTC’s Research Division for financial support.
\end{acknowledgements}



\begin{thebibliography}{190}
	
\bibitem {Wei}{Wei, Q.-H; Su, K.-H.; Durant, S.; and Zhang, X., \textit{Plasmon Resonance of Finite One-Dimensional Au Nanoparticle Chains}, Nano Lett. \tb{4}, 1067 (2004).}

\bibitem {Murray}{Murray, W. A.; Barnes, W. L.  \textit{Plasmonic Materials}, Adv. Mat. \tb{19}, 3771 (2007).}

\bibitem {Carsten}{Jakab A., Rosman C., Khalavka Y., Becker J., Tr\"ügler A., Hohenester U., and S\"onnichsen C., \textit{Highly sensitive plasmonic silver nanorods}, ACS nano {\bf 5}, 6880 (2011).}

\bibitem {Rycenga}{Rycenga M., Cobley C. M., Zeng J., Li W. Y., Moran C. H., Zhang Q., Qin D., and Xia Y. N., \textit{Controlling the Synthesis and Assembly of Silver Nanostructures for Plasmonic Applications}, Chem. Rev. {\bf 111}, 3669 (2011).}

\bibitem {Igarachi} {Igarashi T., Kawai H., Yanagi K., Cuong N. T., Okada S., and Pichler T., \textit{Tuning Localized Transverse Surface Plasmon Resonance in Electricity-Selected Single-Wall Carbon Nanotubes by Electrochemical Doping}, Phys. Rev. Lett. {\bf 114}, 176807 (2015).}

\bibitem{Li} {Lin L., Zapata M., Xiong M, Liu Z. H., Wang S. S., Xu H., Borisov A. G., Gu H. C., Nordlander P., Aizpurua J., and Ye J., \textit{Nanooptics of Plasmonic Nanomatryoshkas: Shrinking the Size of a Core–Shell Junction to Subnanometer}, Opt. Lett. {\bf 15}, 6419 (2015).}


\bibitem{Lerme} {Lerme J., Palpant B., Pr\'evel B., Pellarin M., Treilleux M., Vialle J. L., Perez A., and Broyer M., \textit{Quenching of the Size Effects in Free and Matrix-Embedded Silver Clusters}, Phys. Rev. Lett. {\bf 80}, 5105 (1998).}

\bibitem{Scholl}{Scholl J., Koh A. L., and Dionne J., \textit{Quantum plasmon resonances of individual metallic nanoparticles}, Nature \tb{483}, 421 (2012).}

\bibitem{Garcia}{Garcia de Abajo, F. J.; \textit{Microscopy: Plasmons go quantum}, Nature \tb{483}, 417 (2012).} 

\bibitem{Zuloaga}  {Zuloaga J., Prodan E., and Nordlander P., \textit{Quantum Plasmonics: Optical Properties and Tunability of Metallic Nanorods}, ACS Nano {\bf 4}, 5269 (2010).}

\bibitem{Marinica} {Marinica D. C., Kazansky A. K., Nordlander P., Aizpurua J., and Borisov A. G., \textit{Quantum Plasmonics: Nonlinear Effects in the Field Enhancement of a Plasmonic Nanoparticle Dimer}, Nano Lett. {\bf 12}, 1333 (2012).}

\bibitem{Raza} {Raza S., Stenger N., Kadkhodazadeh S., Fischer S. V., Kostesha N., Jauho A. P., Burrows A., Wubs M., and Mortensen N. A., \textit{Blueshift of the surface plasmon resonance in silver nanoparticles studied with EELS}, Nanophotonics {\bf 2}, 131 (2013).}

\bibitem{Monreal} {Monreal R. C., Tomasz J Antosiewicz T. J., and Apell S. P., \textit{Competition between surface screening and size quantization for surface plasmons in nanoparticles}, New J. Phys. {\bf 15}, 083044 (2013).}

\bibitem{Nordlander}{Talley C. E., Jackson J. B., Oubre C., Grady N. K., Hollars C. W., Lane S. M., Huser T. M., Nordlander P.,and Halas N. J., \textit{Surface-Enhanced Raman Scattering from Individual Au Nanoparticles and Nanoparticle Dimer Substrates}, Nano Lett. {\bf 5}, 1569 (2005).}

\bibitem{Davoyan} {Davoyan A. R., Popov V. V., and Nikitov S. A., \textit{Tailoring Terahertz Near-Field Enhancement via Two-Dimensional Plasmons}, Phys. Rev. Lett. {\bf 108}, 127401 (2012).}

\bibitem{Huang} {Huang Y.,  Ma L. W., Hou M. J., Xieb Z., and  Zhang Z. J., \textit{Gradual plasmon evolution and huge infrared near-field enhancement of metallic bridged nanoparticle dimers}, Phys. Chem. Chem. Phys. {\bf 18}, 2319 (2016).}

\bibitem{zhuzhao} {Zhu J. and Zhao S. M., \textit{A Computational Study of the Giant Local Electric Field Enhancement in Al-Au-Ag Trimetallic Three-Layered Nanoshells}, Plasmonics {\bf 11}, 659 (2016).}

\bibitem{Fung} {Fung C. K. M., Xi N., Lou J. Y., Zhou Z. F., Shanker B., Lai K. W. C., and Chen H. Z., \textit{Quantum Effect in Field Enhancement Using Antenna for Carbon Nanotube Based Infrared Sensors }, Proc. 10th IEEE Conference on Nanotech. {\bf 1}, 458 (2010).}

\bibitem{Singh} {Singh M. R., Schindel D. G., and Hatef A., \textit{Dipole-dipole interaction in a quantum dot and metallic nanorod hybrid system}, Appl. Phys. Lett. {\bf 99}, 181106 (2011).}

\bibitem{Ali} {Hatef A., Sadeghi. S. M., and Singh. M R., \textit{Coherent molecular resonances in quantum dot–metallic nanoparticle systems: coherent self-renormalization
		and structural effects }, Nanotechnology {\bf 23}, 205203 (2012).}

\bibitem{Anton} {Ant\'on M. A., Carre\~no F., Melle S., Calder\'on O. G., and Cabrera-Granado E., \textit{Optical pumping of a single hole spin in a p-doped quantum dot coupled to a metallic nanoparticle}, Phys. Rev. B {\bf 87}, 195303 (2013).}

\bibitem{Ciraci} {Cirac\`i C., Hill R. T., Mock J. J., Urzhumov Y.,  Fernández-Domínguez A. I., Maier S. A., Pendry J. B., Chilkoti A., Smith D. R., \textit{Probing the Ultimate Limits of Plasmonic Enhancement}, Science {\bf 337}, 1072 (2012).}

\bibitem{Zhu} {Zhu W. and  Crozier K. B., \textit{Quantum mechanical limit to plasmonic enhancement as observed by surface-enhanced Raman scattering}, Nature Comm. {\bf 5}, 5228 (2014).} 

\bibitem{Zapata} {Zapata M., Camacho A. S.,	Borisov A. G., and Aizpurua J., \textit{Quantum effects in the optical response of extended plasmonic gaps: validation of the quantum corrected model in core-shell nanomatryushkas}, Opt. Express {\bf 23}, 8134 (2015).}

\bibitem{Chang}  {Chang H. W. and Mu S. Y., \textit{Semi-analytical solutions of the 3D homogeneous Helmholtz equation by the method of connected local fields}, Prog. Electromagn. Res. {\bf 142}, 159 (2013).}

\bibitem{Arnaud}  {Deraemaeker A., Babu\v{s}ka I., and Bouillard P., \textit{Dispersion and pollution of the FEM solution for the Helmholtz equation in one, two and three dimensions}, Int. J. Numer. Meth. Eng. {\bf 46}, 471 (1999).}

\bibitem{Misic} {Micic M., Klymyshyn N., Suh Y. D., and Lu H. P., \textit{Finite Element Method Simulation of the Field Distribution for AFM Tip-Enhanced Surface-Enhanced Raman Scanning Microscopy}, J. Phys. Chem. B {\bf 107}, 1574 (2003).}

\bibitem{Santana} {Ram\'irez H. Y. and Santana A., \textit{Two interacting electrons confined in a 3D parabolic cylindrically symmetric potential, in presence of axial magnetic field: A finite element approach}, Comput. Phys. Comm. {\bf 183}, 1654 (2012).}

\bibitem{forou} {Forouzeshfard M. R. and Hosseini Farzad M., \textit{Electromagnetic Wave Propagation Through Two Coaxial Transformation-based Cylindrical Media}, Plasmonics {\bf 10}, 1345 (2015).}

\bibitem{Jacak1} {Kluczyk K., and Jacak W., \textit{Damping-induced size effect in surface plasmon resonance in metallic nano-particles: Comparison of RPA microscopic model with numerical finite element simulation (COMSOL) and Mie approach}, J Quant. Spectrosc. Rad. {\bf 168}, 78 (2016).}

\bibitem{comsol} http://www.comsol.com

\bibitem{drudemodel} Maier S. A., \textit{Plasmonics: Fundamentals and Applications}, New York - USA; Springer, (2007), Chapter 1.

\bibitem{Kelly} {Kelly K. L., Coronado E., Zhao L. L., and Schatz G. C., \textit{The Optical Properties of Metal Nanoparticles: The Influence of Size, Shape, and Dielectric Environment}, J. Phys. Chem. B {\bf 107}, 668 (2003).}

\bibitem{Grady} {Grady N. K., Halas N. J., Nordlander P., \textit{Influence of dielectric function properties on the optical response of plasmon resonant metallic nanoparticles}, Chem. Phys. Lett. {\bf 399}, 167 (2004).}

\bibitem{Noguez} {Noguez C. \textit{Surface Plasmons on Metal Nanoparticles: The Influence of Shape and Physical
		Environment}, J. Phys. Chem. C {\bf 111}, 3806 (2007).}

\bibitem{Esteban} {Esteban R.,	Borisov A. G.,	Nordlander P., and  Aizpurua J., \textit{Bridging quantum and classical plasmonics with a quantum-corrected model}, Nature Comm. {\bf 3}, 825 (2012).} 

\bibitem{Muneton} {Muñetón Arboleda D., Santillán J. M. J., Mendoza Herrera L. J., Muraca D., Schinca D. C., and Scaffardi L. B., \textit{Size-dependent complex dielectric function of Ni, Mo, W, Pb, Zn and Na nanoparticles. Application to sizing}, J. Phys. D {\bf 49}, 075302 (2016).}

\bibitem{Genzel}{Genzel L., Martin T. P., and Kreibig U., \textit{Dielectric function and plasma resonances of small metal particles}, Z. Phys. B \tb{21}, 339, (1975).}

\bibitem{Pccp} {Ram\'irez H. Y., Fl\'orez J., and Camacho A. S., \textit{Efficient control of coulomb enhanced second harmonic generation from excitonic transitions in quantum dot ensembles}, Phys. Chem. Chem. Phys. {\bf 17}, 23938 (2015).}

\bibitem{Arfken}{Mathematical Methods for Physicists, G. B. Arfken and H. J. Weber. 5th Ed., Harcourt/Academic Press, 2001.}

\bibitem{JandC}{Johnson and Christy, \textit{Optical Constants of the Noble Metals}, Phys. Rev. B {\bf 6}, 4370 (1972).} 

\bibitem{charle} {Charl\'e K. P., Schulze W., and Winter B., \textit{The size dependent shift of the surface plasmon absorption band of small spherical metal particles}, Z. Physik D {\bf 12}, 471 (1989).}

\bibitem{Negre}  {Negre F. A., Perassi E. M., Coronado E. A. and S\'anchez C. G., \textit{Quantum dynamical simulations of local field enhancement in metal nanoparticles}, J. Phys.: Condens. Matter {\bf 25}, 125304 (2013).}

\bibitem{Blaber} {Blaber M. G., Arnold M. D., and Ford M. J., \textit{Search for the Ideal Plasmonic Nanoshell: The Effects of Surface Scattering and Alternatives to Gold and Silver}, J. Phys. Chem. C {\bf 113}, 3041 (2009).} 

\bibitem{Jacak}{Jacak W. A., \textit{Lorentz Friction for Surface Plasmons in Metallic Nanospheres}, J. Phys. Chem. C {\bf 119}, 6749 (2015).}

\bibitem{Marzan}{Liz-Marz\'an L. M., \textit{Tailoring Surface Plasmons through the Morphology and Assembly of Metal Nanoparticles}, Langmuir {\bf 22}, 32 (2006).}

\bibitem{Zhang} {Zhang S. P., Bao K., Halas N. J., Xu H. X., and Nordlander P., \textit{Substrate-Induced Fano Resonances of a Plasmonic Nanocube: A Route to Increased-Sensitivity Localized Surface Plasmon Resonance Sensors Revealed}, Nano Lett. {\bf 11}, 1657 (2011).}

\bibitem{ZhangY} {Zhang Y., Jia T. Q., Zhang S. A., Feng D. H.,	and Xu Z. Z., \textit{Dipole, quadrupole and octupole plasmon resonance modes in non-concentric nanocrescent/nanodisk structure: local field	enhancement in the visible and near infrared regions}, Opt. Express {\bf 20}, 2924 (2012).}

\bibitem{Heer} {de Heer W. A., \textit{The physics of simple metal clusters: experimental aspects and simple models}, Rev. Mod. Phys. {\bf 65}, 611 (1993).}

\bibitem{Brack} {Brack M., \textit{The physics of simple metal clusters: self-consistent jellium model and semiclassical approaches}, Rev. Mod. Phys. {\bf 65}, 677 (1993).} 

\end{thebibliography}


\end{document}